# Growth-induced stability in modified SLE curve

Yusuke Shibasaki*
Nihon University, Tokyo

*yshibasaki98@gmail.com

Abstract

In this study, the non-equilibrium free energy corresponding to the curve generated by a modified stochastic Loewner evolution (SLE), which is driven by the Langevin equation, is theoretically investigated. Under certain conditions, we prove that the time derivative of the (generalized) free energy expressed by Kullback-Leibler divergence between the probability distributions of the curve and driving function has a positive value, indicating the negativity of Gibbs entropy production. In addition, it was implied that, in a certain restriction, the free energy can be expressed as a function of a Lyapunov-type exponent of the driving function. These results show a dissipative nature of conformal dynamics, and indicate the growth-induced stability of the modified SLE curve.

**I. INTRODUCTION**

The stochastic Loewner evolution (SLE) suggested by Schramm has been shown to provide a scaling limit of the random geometry appearing in the two-dimensional (2D) statistical mechanics models.[1, 2] The SLE is a growth process of a random curve in the complex plane determined by Loewner differential equation driven by the Winer process. In the physics context, the static geometry generated by this model is frequently discussed. However, the dynamical properties of the curve also is worthwhile examining because the 2D self-organization phenomena described by conformal map systems, e.g., the diffusion-limited aggregation (DLA),[3] Laplacian growth (LG)[4] or Laplacian random walk (LRW),[5] have deep connections with the Loewner equation and the SLE.[6] Although the basic SLE theory assumes the conformal invariance, which particularly holds for the critical phenomena, its broader applications to the real 2D growth systems requires an extension of the standard SLE. In Ref. 7, the authors suggested a modified SLE driven by Langevin dynamics as a possible form of such extended models. This modification of the driving function not only expanded the descriptive ability of the framework but also resulted in a one-to-one correspondence between the non-equilibrium system (curve growth) and the equilibrium system (driving function). In this modified SLE, the irreversible and non-equilibrium properties of the curve growth were measured by the relative entropy between the curve and driving function, and it can be calculated as a function of the conformal maps parametrized by time parameter in the Loewner equation. In Ref. 7, the authors referred to it as the *relative Loewner entropy*, and indicated that it closely relates to the Lyapunov-type exponent of the conformal maps.

In the context of statistical mechanics, the relative entropy, (i.e., Kullback–Leibler (KL) divergence) is interpreted as being related to the free energy, and *H-function* of the system.[8-10] In most of the equilibrium and non-equilibrium systems, the time derivative of the free energy assumes a negative value, which indicates the positivity of the entropy production inside the system.[8-11] However, for the systems involving dissipations of the energy, such as self-organization phenomenon in living biological systems or non-equilibrium physical systems, the negativity of the total entropy production of the system is often reported. Such examples include negative production of total entropy in the photosynthesis systems[12-14] and non-Markovian dynamics in physical systems. This implies that, for the processes in non-equilibrium systems, there is a possibility that the time derivative of the free energy assumes a positive value, i.e., negative total entropy production. However, there are few mathematical models that inherently describe this principle in the statistical mechanics context. It should be noted that this argument does not necessarily indicate the violation of the second law because most of the above-mentioned dissipative systems are assumed to be open systems, which allow matter transmissions with the external environments.

In this study, the mathematical evidence for the above discussion is provided by demonstrating the modified SLE as an example system whose time derivative of free energy assumes positive value. The present results are expected to provide a possible theoretical interpretation for the 2D morphogenetic or self-organizing phenomena, which can be dealt with in the context of non-equilibrium statistical physics. The remainder of this paper is organized as follows. In Sec. II, the model and definitions of the quantities used in this study are introduced. In Sec. III, the positivity of the time derivative of the free energy using the relative entropy is proven; this is the main result of this study. In Sec. IV, the stationarity of the modified SLE trajectories is proven to demonstrate the equivalence between the Lyapunov-type exponent of the conformal maps and the free energy function. In Sec. V, the conclusions based on the present result are drawn.

**II. DEFINITION AND FORMULA**

**A. Modified SLE**

Let $\xi_t \in \mathbb{R}$ be a continuous function parametrized by time $t \in [0, \infty)$, whose evolution is governed by the stochastic differential equation (SDE), i.e., the Langevin equation,

$$d\xi_t = \mu(\xi_t)dt + \sqrt{\kappa}dB_t. \tag{1}$$

Here, $B_t$ is a standard Brownian motion on $t \in [0, \infty)$, and $\kappa \in \mathbb{R}$ is a constant diffusivity parameter. Let us denote the smallest $\sigma$-field generated by $B_t$ up to $t$ as $\mathcal{F}_t \coloneqq \sigma(B_s: 0 \leq s \leq t)$, which is also called filtration. We assume that the drift term $\mu(\xi_t)$ is a $\mathcal{F}_t$-measurable continuous function with the potential function $V(\xi_t)$ satisfying,

$$\mu(\xi_t) = -\frac{\partial V(\xi_t)}{\partial \xi}. \tag{2}$$

We assume that $V(\xi_t)$ is a spatially symmetric function satisfying $V(-\xi_t) = -V(\xi_t)$. The equilibrium distribution of the driving function is expressed as

$$p_{eq}(\xi) = Z_{nor}^{-1} \exp\left[-2V(\xi_t)/\kappa\right], \tag{3}$$

where $Z_{nor}$ is a normalization constant. We consider the chordal Loewner evolution driven by the stochastic process expressed by Eq. (1). Let us denote the curve starting at the origin in the upper half complex plane $\mathbb{H}$ as $\gamma[0, t]$. Let $g_t(z)$ with $z \in \mathbb{H}$, be the conformal map from $\mathbb{H} \setminus \gamma[0, t]$ to $\mathbb{H}$, satisfying[2, 7]

$$g_t(z) = z + \frac{2t}{z} + O(|z|^{-2}), \quad as \quad z \to \infty. \tag{4}$$

The chordal Loewner evolution driven by the shifted driving function $\xi_t - \xi_0$ is expressed as the following differential equation of the map $g_t(z)$:[7]

$$\frac{\partial g_t(z)}{\partial t} = \frac{2}{g_t(z) - (\xi_t - \xi_0)}, \tag{5}$$

where $g_0(z) = z$. Let us denote the tip of $\gamma[0, t]$ as $z_t \in \mathbb{H}$. By considering the condition that the dominator of the right-hand side of Eq. (5) equals to zero, and implying that the curve is not analytic, the following relation is derived:[7]

$$g_t(z_t) = \xi_t - \xi_0. \tag{6}$$

Although we are concerned with the time evolution of the tip of the curve $z_t$, the Loewner equation expressed as Eq. (5) describes the time evolution of $g_t(z)$. For the purpose of this study, it is useful to discuss the backward flow of the Loewner evolution[2, 15, 16] as described in the subsequent subsection.

**B. Backward Loewner evolution**

The time evolution of the tip of the SLE curve is often discussed by considering the time-reversed conformal map $g_{-t}$. Considering the following the backward Loewner evolution,[2, 15]

$$\frac{\partial f_t(w)}{\partial t} = -\frac{2}{f_t(w) - (\xi_t - \xi_0)}, \tag{7}$$

where $f_0(w) = w \in \mathbb{H}$, and $f_t := g_t^{-1}$, we obtain the following lemma. (an analog of the Lemma 3.1 in Ref. 2).

**Lemma 1.** *Let $\xi_t$ have an equilibrium (stationary) distribution, expressed by Eq. (3) from the initial conditions. Then, for all fixed $t \in \mathbb{R}$, the map $w \mapsto f_t(w)$ has the same distribution as the map $w \mapsto g_t^{-1}(w - (\xi_t - \xi_0)) + (\xi_t - \xi_0)$.*

*Proof.* From Eqs.(2) and (3), if $\xi_t$ has an equilibrium probability distribution function and $V(\xi_t) = -V(\xi_t)$, $\xi_t$ becomes a reversible function in the sense that $\xi_t$, $\xi_{-t}$, and $-\xi_{-t}$ have the same distribution. Then, for a fixed $t_1 \in \mathbb{R}$, let $\hat{\xi}_{t_1}(t) = \xi_{t_1+t} - \xi_{t_1}$. From the stationary increment property of $\xi_t$, $\hat{\xi}_{t_1}: \mathbb{R} \to \mathbb{R}$ has the same distribution as $\xi_t: \mathbb{R} \to \mathbb{R}$. Considering the process

$$\hat{g}_t(w) := g_{t_1+t} \circ g_{t_1}^{-1}\left(w - (\xi_{t_1} - \xi_0)\right) + (\xi_{t_1} - \xi_0), \qquad \hat{g}_0(w) = w, \tag{8}$$

we obtain

$$\hat{g}_{-t_1}(w) = g_{t_1}^{-1}(w - (\xi_{t_1} - \xi_0)) + (\xi_{t_1} - \xi_0). \tag{9}$$

From the translation invariance of $\xi_t$ and Eq. (8), we obtain the following:

$$\frac{\partial \hat{g}_t(w)}{\partial t} = \frac{2}{\hat{g}_t(w) + \xi_{t_1} - \xi_{t_1+t} + \xi_0} = \frac{2}{\hat{g}_t(w) - (\hat{\xi}_{t_1}(t) - \xi_0)}. \tag{10}$$

Since $\hat{g}_{-t_1}(w) := f_t\left(w + (\xi_{t_1} - \xi_0)\right) - \xi_{t_1}$, from Eqs. (7) and (10), it can be observed that $\hat{g}_{-t}(w)$ has the same distribution as $f_t(w)$. Consequently, the lemma is obtained from Eq. (9).

From the above lemma, the Langevin equation for the time evolution of the tips of the backward evolution, defined as $\hat{z}_t := f_t(w) - (\xi_t - \xi_0)$, is derived as follows.[15]

$$\frac{\partial \hat{z}_t}{\partial t} = -\frac{2}{\hat{z}_t} - \frac{\partial \xi_t}{\partial t}. \tag{11}$$

Note that $z_t$ and $\hat{z}_t$ have the same probability distributions. Separating the real and imaginary parts of $\hat{z}_t = x_t + i y_t$, and using Eq. (1), we obtain,[7]

$$\frac{dx_t}{dt} = -\frac{2x_t}{x_t^2 + y_t^2} - \mu(\xi_t) - \sqrt{\kappa}\frac{dB_t}{dt}, \tag{12}$$

$$\frac{dy_t}{dt} = \frac{2y_t}{x_t^2 + y_t^2}. \tag{13}$$

The Fokker-Planck equation associated with Eqs. (12) and (13) can be expressed as[7]

$$\frac{\partial p(x,y,t)}{\partial t} = \left\{\frac{\kappa}{2}\frac{\partial^2}{\partial x^2} + \frac{\partial}{\partial x}\left[\frac{2x}{x^2+y^2} + \mu(\xi)\right] - \frac{\partial}{\partial y}\frac{2y}{x^2+y^2}\right\} p(x,y,t), \tag{14}$$

where $p(x,y,t)$ is the probability measure for $\hat{z}_t$, satisfying $p(x,y,t) = \langle \delta(x_t - x)\delta(y_t - y)\rangle$. To ensure Eq. (14) depicts the behavior of the tip $z_t$ of the modified SLE, the initial condition is set as $x_0 = 0$ and $y_0 = \varepsilon$, where $\varepsilon$ is an infinitesimal positive constant.[15]

### C. Gibbs entropy, free energy and Lyapunov exponent

Using the above equations describing the time evolution of the tip of the modified SLE curve, we define the several quantities, which we use for our discussion. Firstly, the Gibbs entropy of the tip of the modified SLE curve is defined as the following.

**Definition 1:** (Gibbs entropy) *The Gibbs entropy of the tip of the curve of the modified SLE described by Eq. (5) is defined as*

$$S(t) := -k_B \iint p(x,y,t) \ln p(x,y,t)\, dx dy, \tag{15}$$

*where $k_B \in \mathbb{R}$ is a positive (Boltzmann) constant.*

In the statistical physical sense, $S(t)$ is the entropy, whose ensemble average is calculated over every possible configuration of the tip of the curve. The above-defined Gibbs entropy is used for both the equilibrium and non-equilibrium states owing to its time dependency. Subsequently, we define the relative entropy (i.e., KL divergence) between the probability measure of the tip of the curve $z_t$ and the driving function $\xi_t$.

**Definition 2**: (KL divergence) *The KL divergence between the tip of the curve $z_t$ and the driving function $\xi_t$ is defined as*

$$D(z \parallel \xi) := \iint p(x, y, t) \ln \frac{p(x, y, t)}{p(\xi)} dx dy, \qquad (16)$$

*and the following is denoted*

$$d(z \parallel \xi) := \ln \frac{p(x, y, t)}{p(\xi)}. \qquad (17)$$

In this study, we consider the condition that $p(\xi)$ is a stationary (equilibrium) distribution described by Eq. (3), that is, $p(\xi) = p_{eq}(\xi)$, unless otherwise indicated. From the one-to-one correspondence of the curve and driving function, $d(z \parallel \xi)$ is uniquely determined for an arbitrary trajectory of the curve. Another expression of $d(z \parallel \xi)$ is derived as follows.[7] Considering the transformation $z = g_t^{-1}(w)$ with $w \in \mathbb{H}$, we immediately obtain

$$p(z, t) = p(w, t) \left| \frac{dg_t(z)}{dz} \right|^2. \qquad (18)$$

Using this equation and $\xi_t - \xi_0 = \lim_{z \to z_t} g_t(z)$ from Eq. (6), the following relation is derived:

$$d(z \parallel \xi) = \ln \left| \frac{dg_t(z_t)}{dz} \right|^2. \qquad (19)$$

We shall use this expression in subsequent discussions. The physical meaning of the relative entropy has been discussed in several studies.[8-10] Because it measures the divergence between two different probability distribution function in a manner based on the Gibbs entropy formula, it is interpreted as the free energy difference between the two configurations. Using the relative entropy, we define the non-equilibrium free energy function for the tip of the curve trajectory. The equilibrium distribution of the driving function in Eq. (3) is rewritten as $p_{eq}(\xi) = Z_{nor}^{-1} \exp[-\beta V(\xi_t)]$, where $\beta :=$ $1/k_B T = 2/\kappa$, and $T$ is the temperature in the physical sense. Notably, the equilibrium distribution of the curve trajectory is same as that of the driving function by considering the limit of $\hat{z}_t \to \infty$ in Eq. (11). It implies that, if the driving function is established in an equilibrium state, $D(z \parallel \xi)$ in Eq. (16) measures the divergence between the probability of the curve trajectory at time $t$ and that of its possible equilibrium state $p_{eq}(x,y)$, [i.e., $d(z \parallel \xi) \simeq \ln \frac{p(x,y,t)}{p_{eq}(x,y)}$]. Considering that the curve trajectory is a system that exchanges the energy with the external environment whose potential is represented by $V(\xi_t)$, the following definition of the (non-equilibrium) Helmholtz free energy function is physically consistent.[9-11]

**Definition 3.** *The non-equilibrium free energy function for the tips of the modified SLE curve is defined as follows:*

$$F(t) := U - TS(t), \qquad U := \langle V(\xi_t) \rangle. \tag{20}$$

Here, the brackets denote the ensemble average of the potential $V(\xi_t)$ with respect to the equilibrium distribution of the driving function $\xi_t$. Noting that the free energy for the equilibrium condition for both the curve and driving function is expressed as $F_{eq} = -\beta^{-1} \ln Z_{nor}$, the following expression of the free energy function using the relative entropy is plausible in the physical sense.

$$F(t) = F_{eq} + k_B T D(z \parallel \xi), \tag{21}$$

where $k_B, T$, and $F_{eq} \in \mathbb{R}$ are constant over time. In the following discussion, we consider the time derivative of $F(t)$, which is expressed as

$$\frac{dF(t)}{dt} = k_B T \frac{dD(z \parallel \xi)}{dt}. \tag{22}$$

As discussed above, for isolated systems or many the non-equilibrium systems in the statistical physics model, $dF(t)/dt$ assumes a negative value or zero. On the contrary, in Theorem 1, we demonstrate that there are conditions in which the modified SLE curves have $dF(t)/dt > 0$. Furthermore, it shall be demonstrated that $F(t)$ converges to a constant under certain conditions. For these discussions, the ergodicity of the tip of the curve trajectory is required, and it is convenient to define the (modified) Lyapunov exponent of the driving function. Therefore, we define the complex Lyapunov exponent of the driving function as follows:

**Definition 4.** *Let $\{\omega_i\}_{i=0}^n$ be the time dependent transformation of $\xi_t \in \mathbb{C}$, satisfying $\xi_n = \omega_n \circ \omega_{n-1} \circ \cdots \circ \omega_0(\xi_0)$. The Lyapunov exponent of the driving function $\lambda^\xi$ is defined as*

$$\lambda^\xi := \lim_{n \to \infty} \lim_{\delta z_0 \to 0} \frac{1}{n} \ln \left| \frac{\delta \xi_n}{\delta z_0} \right|, \tag{23}$$

*where $\delta \xi_n = \omega_n \circ \omega_{n-1} \circ \cdots \circ \omega_0(\xi_0 + \delta z_0) - \omega_n \circ \omega_{n-1} \circ \cdots \circ \omega_0(\xi_0)$ and $\delta z_0$ is a complex-numbered infinitesimal separation.*

The physical meaning of $\lambda^\xi$ is as follows. In the Loewner evolution, the driving function is defined as that evolving on the real axis. The Lyapunov exponent for $\xi_t$ in the usual sense measures the separation rate of the different initial conditions whose separation is an infinitesimal real-numbered distance $\delta \xi_0 \in \mathbb{R}$. Contrarily, the complex Lyapunov exponent $\lambda^\xi$ is defined as that measures the separation rate of the infinitesimal complex separation $\delta z_0 \in \mathbb{C}$. Thus, the orbit of $\xi_t$ perturbated by $\delta z_0$ moves on the complex plane $\mathbb{C}$. In Theorem 3, it is shown that $\lambda^\xi$ can be described in terms of $D(z \parallel \xi)$, which is an important quantity for our discussion.

### III. TIME DERIVATIVE OF FREE ENERGY FUNCTION

The free energy describing non-equilibrium systems has been studied in relation to the *H*-theorem and the second law of thermodynamics.[8-11] For the generalized Fokker-Planck equations, the negativity of its time derivative of the free energy $dF(t)/dt \leq 0$ has been proven with some restrictions.[9-11] The following shows the opposite results obtained by the present model, which suggests the negativity of the entropy production. The following theorem is one of our main results.

**Theorem 1.** *If* $\left|\frac{-2x_t}{x_t^2+y_t^2}\right| < |\mu(\xi_t)|$ *and* $y_t > |x_t|$, *the time derivative of the free energy corresponding to the trajectory of the tip of the modified SLE curve is positive, i.e.,*

$$\frac{dF(t)}{dt} > 0 \tag{24}$$

*holds true for sufficiently large* $t$.

*Proof.* Using Eq. (14) and integrating by parts, the time derivative of $D(z \| \xi)$ is obtained as follows:

$$\begin{aligned}
\frac{dD(z \| \xi)}{dt} &= \iint \left[ \ln \frac{p(x,y,t)}{p_{eq}(\xi)} \frac{d}{dt} p(x,y,t) + p(x,y,t) \frac{d}{dt} \ln \frac{p(x,y,t)}{p_{eq}(\xi)} \right] dxdy \\
&= \iint \left[ 1 + \ln \frac{p(x,y,t)}{p_{eq}(\xi)} \right] \left\{ \frac{\kappa}{2} \frac{\partial^2}{\partial x^2} + \frac{\partial}{\partial x} \left[ \frac{2x}{x^2+y^2} + \mu(\xi) \right] - \frac{\partial}{\partial y} \frac{2y}{x^2+y^2} \right\} p(x,y,t) \, dxdy \\
&= -\iint \frac{\kappa}{2p(x,y,t)} \left[ \frac{\partial}{\partial x} p(x,y,t) \right]^2 + \mu(\xi) \frac{\partial}{\partial x} p(x,y,t) - p(x,y,t) \left\{ \frac{\partial}{\partial x} \frac{2x}{x^2+y^2} - \frac{\partial}{\partial y} \frac{2y}{x^2+y^2} \right\} dxdy \\
&= -\iint \left\{ \frac{\kappa}{2p(x,y,t)} \left[ \frac{\partial}{\partial x} p(x,y,t) \right]^2 \left[ 1 + \frac{2\mu(\xi)}{\kappa \frac{\partial}{\partial x} \ln p(x,y,t)} \right] - p(x,y,t) \left[ \frac{4(y^2-x^2)}{(x^2+y^2)^2} \right] \right\} dxdy.
\end{aligned}$$

$$\tag{25}$$

Here, we drop the boundary terms $p(x,y,t)$ to zero in the limits of $x \to \pm\infty$ and $y \to \pm\infty$. To evaluate the sign of the right-hand side of Eq. (25), we set

$$A(x,y,\xi,t) = 1 + \frac{2\mu(\xi)}{\kappa \frac{\partial}{\partial x} \ln p(x,y,t)}, \tag{26}$$

and

$$B(x,y) = \frac{4(y^2-x^2)}{(x^2+y^2)^2}. \tag{27}$$

First, we evaluate the sign of $A(x,y,\xi,t)$. Note that for a fixed $y$,

$$\frac{\partial}{\partial x} p(x,y,t) = p(y,t) \frac{\partial}{\partial x} p(x,t). \tag{28}$$

From Eq. (12), we obtain the stationary distribution of $x_t$ as

$$p(x) = {Z'_{nor}}^{-1} \exp\left[-\frac{2}{\kappa} \int \left( \frac{2x_t}{x_t^2+y_t^2} + \mu(\xi_t) \right) dx \right]. \tag{29}$$

where $Z'_{nor}$ is a normalization constant. Considering the logarithm of Eq. (29), we obtain

$$\ln p(x) = \ln \frac{1}{Z'_{nor}} - \frac{2}{\kappa} \int \left( \frac{2x_t}{x_t^2 + y_t^2} + \mu(\xi_t) \right) dx. \tag{30}$$

From Eqs. (26) and (30), we observe that

$$A(x, y, \xi, t) = 1 + \frac{\mu(\xi_t)}{-\frac{2x_t}{x_t^2 + y_t^2} - \mu(\xi_t)}. \tag{31}$$

If $\left| \frac{-2x_t}{x_t^2 + y_t^2} \right| < |\mu(\xi_t)|$, we obtain

$$A(x, y, \xi, t) < 0. \tag{32}$$

Moreover, if $y_t > |x_t|$, we obtain

$$B(x, y) > 0. \tag{33}$$

Substituting Eqs. (32) and (33) into Eq. (25), we obtain

$$\frac{dD(z \parallel \xi)}{dt} > 0, \tag{34}$$

provided that $\left| \frac{-2x_t}{x_t^2 + y_t^2} \right| < |\mu(\xi_t)|$ and $y_t > |x_t|$. From Eq. (22), we observe that Eq. (34) indicates that $dF(t)/dt > 0$.

**Remark 1.** *For a sufficiently large $t$, the condition $y_t > |x_t|$ holds true because $y_t$ is a monotonically increasing function from Eq. (13), while $p(x)$ approaches to a stationary distribution such that $|x_t| < C$, where $C$ is a certain positive constant.*

**Corollary 1.** *When the above theorem holds true, the Gibbs entropy $S(t)$ becomes a decreasing function in time.*

*Proof. It follows from Eqs. (20) and (24).*

## IV. STATIONARITY AND LONG-TIME LIMIT OF FREE ENERGY

Examining the existence of the invariant measure for the tip of the curve is essential in discussing the convergence of $F(t)$ in the long-time limit. For the standard SLE, the ergodicity of the tip of the curve is shown in Ref. 17. We demonstrate that the present model satisfies the stationarity, which implies ergodicity, with some restrictions on $\mu(\xi_t)$. Let us consider the following Bessel-type process $Z_t$ associated with the present modified SLE.[18, 19]

$$dZ_t = dB_t + \frac{\sqrt{\delta-1}}{2}\mu(\xi_t)dt + \frac{\delta-1}{2}\frac{dt}{Z_t}. \tag{35}$$

If we set $\tilde{g}_t(z) = Z_t - \frac{\sqrt{\delta-1}}{2}\int_0^t \mu(\xi_s)ds - B_t + \frac{\sqrt{\delta-1}}{2}\xi_0$, where $\tilde{g}_t(z) := (1/\sqrt{\kappa})g_t(z)$, we obtain the Loewner equation in Eq. (5), under the transformation $(\xi_t - \xi_0) \to -(\xi_t - \xi_0)$ and

$$\delta = 1 + \frac{4}{\kappa}. \tag{36}$$

The phases and other important properties of the SLE curve are shown by the stochastic properties of the Bessel process. Because $z_t$, $\hat{z}_t$, and $Z_t$ have the same distribution, we can observe the long-time behavior and convergence of $p(x,y,t)$ by analyzing Eq. (35). Separating $Z_t = X_t^{\tilde{x}} + iY_t^{\tilde{y}}$ in Eq. (35) into the real and imaginary parts, we obtain the following equations:[19]

$$dX_t^{\tilde{x}} = dB_t + \frac{\sqrt{\delta-1}}{2}\mu(\xi_t)dt + \frac{\delta-1}{2}\frac{X_t^{\tilde{x}}}{\left(X_t^{\tilde{x}}\right)^2 + \left(Y_t^{\tilde{y}}\right)^2}dt, \tag{37}$$

$$dY_t^{\tilde{y}} = -\frac{\delta-1}{2}\frac{Y_t^{\tilde{y}}}{\left(X_t^{\tilde{x}}\right)^2 + \left(Y_t^{\tilde{y}}\right)^2}dt. \tag{38}$$

If $\delta > 1$, the right-hand side of Eq. (38) tends to zero, i.e., $Y_t^{\tilde{y}} \to 0$ as $t \to \infty$. This means that we should analyze particularly the Bessel-type process in Eq. (37) with $Y_t^{\tilde{y}} = 0$ to observe the long-time limit behavior. Let us consider the Bessel flow starting at $X_0^{\tilde{x}} = \tilde{x} > 0$. In the standard SLE driven by the Wiener process, it is well-known that a critical dimension of the Bessel flow is $\delta = \delta_c = 2$, which is a dimension that determines the recurrence or transience of the Bessel flow.[18, 19] For the modified SLE as well, we obtain the following lemma.

**Lemma 2.** (i) *If $\delta > 2$, the Bessel flow of the modified SLE is transient, i.e., $\lim_{t\to\infty} X_t^{\tilde{x}} = \infty$ for all $\tilde{x} > 0$, with probability 1.*

(ii) *If $\delta = 2$, then $\inf_{t>0} X_t^{\tilde{x}} = 0$ for all $\tilde{x} > 0$, with probability 1.*

(iii) *If $1 < \delta < 2$, the Bessel flow of the modified SLE is recurrent, i.e., $T^{\tilde{x}} < \infty$ for all $\tilde{x} > 0$, with*

probability 1, where $T^{\tilde{x}}$ is the first visiting time at $X_t^{\tilde{x}} = 0$.

The proof of this lemma and the definitions of the variables are provided in the appendix. The stationarity of the tip of the curve is related to the behavior of the Bessel-type process. We obtain the following Lemma:

**Lemma 3.** *Let $p(t, \tilde{x}'|\tilde{x})$ be the transition probability density of the Bessel-type process from $\tilde{x}$ to $\tilde{x}'$ The trajectory of the tips of modified SLE has a stationary probability density, in the sense that it satisfies*

$$|p(t, \tilde{x}'|\tilde{x}) - p(\tilde{x}')| \leq Ce^{-\lambda t}, \tag{39}$$

*if*

$$\frac{1}{\tilde{x}} \geq -\frac{\sqrt{\kappa}}{2}\mu(\xi_0) \quad \text{for} \quad X_0^{\tilde{x}} = \tilde{x} > 0,$$

$$\frac{1}{\tilde{x}} \leq \frac{\sqrt{\kappa}}{2}\mu(\xi_0) \quad \text{for} \quad X_0^{\tilde{x}} = \tilde{x} < 0. \tag{40}$$

*Proof.* First, we consider the process starting at $X_0^{\tilde{x}} = \tilde{x} > 0$. The Bessel-type process in Eq. (37) with $Y_t^{\tilde{y}} = 0$ yields the following backward Kolmogorov equation (BKE):

$$\frac{\partial}{\partial t}p(t,\tilde{x}'|\tilde{x}) = \frac{1}{2}\frac{\partial^2}{\partial \tilde{x}^2}p(t,\tilde{x}'|\tilde{x}) + \left[\frac{\delta-1}{2\tilde{x}} + \frac{\sqrt{\delta-1}}{2}\mu(\xi_0)\right]\frac{\partial}{\partial \tilde{x}}p(t,\tilde{x}'|\tilde{x}), \tag{41}$$

with the initial condition $p(0, \tilde{x}'|\tilde{x}) = \delta(\tilde{x}' - \tilde{x})$. Let $\psi_n(t, \tilde{x}'|\tilde{x})$ be the *n*th eigenfunction of Eq. (41). Defining the generator of the Bessel-type process as $\mathcal{L} := \frac{1}{2}\frac{\partial^2}{\partial \tilde{x}^2} + \left[\frac{\delta-1}{2\tilde{x}} + \frac{\sqrt{\delta-1}}{2}\mu(\xi_0)\right]\frac{\partial}{\partial \tilde{x}}$, the following is obtained:

$$\mathcal{L}\psi_n(t, \tilde{x}'|\tilde{x}) = -\lambda_n \psi_n(t, \tilde{x}'|\tilde{x}), \tag{42}$$

where $\lambda_n \in \mathbb{R}$ is *n*th eigenvalue of Eq. (41). The eigenvalues $\lambda_n$ take zero or greater than zero, and can be sorted in increasing order as $0 \leq \lambda_0 < \lambda_1 < \cdots$. Subsequently, $p(t, \tilde{x}'|\tilde{x})$ is expressed as[20]

$$p(t, \tilde{x}'|\tilde{x}) = e^{\mathcal{L}t}\delta(\tilde{x}' - \tilde{x}) = \sum_{n=0}^{\infty} \alpha_n \psi_n(t, \tilde{x}')\psi_n(0, \tilde{x})e^{-\lambda_n t}, \tag{43}$$

where $\alpha_n$ is an *n*-dependent constant. Under the assumption that $p(t, \tilde{x}'|\tilde{x})$ is a stationary state denoted as $p_{st}(t, \tilde{x}'|\tilde{x})$, from Eq. (41) we derive the following ordinal differential equation

$$\frac{\partial^2}{\partial \tilde{x}^2} p_{st}(t, \tilde{x}'|\tilde{x}) + \left[\frac{\delta-1}{\tilde{x}} + \sqrt{\delta-1}\mu(\xi_0)\right]\frac{\partial}{\partial \tilde{x}} p_{st}(t, \tilde{x}'|\tilde{x}) = 0. \tag{44}$$

To find the stable condition of $p(t, \tilde{x}'|\tilde{x})$ with respect to $\tilde{x}$, assuming the stationary probability distribution has a form of $p_{st}(t, \tilde{x}'|\tilde{x}) = C_0 e^{-r_0 \tilde{x}} + C_1 e^{-r_1 \tilde{x}}$, where $C_n$ and $r_n$ ($n=0,1$) are constant parameters, we obtain

$$r_0 = 0 \quad \text{and} \quad r_1 = \frac{\delta-1}{\tilde{x}} + \sqrt{\delta-1}\mu(\xi_0). \tag{45}$$

The existence of $r_0 = 0$ suggests that $p(t, \tilde{x}'|\tilde{x})$ has a stationary solution with respect to $\tilde{x}$. Furthermore, $p_{st}(t, \tilde{x}'|\tilde{x})$ is stable with respect to the initial condition $\tilde{x}$ if $r_1$ in Eq. (45) assumes a positive value. This condition is expressed as the following:

$$\frac{1}{\tilde{x}} \geq -\frac{\sqrt{\kappa}}{2}\mu(\xi_0). \tag{46}$$

Here, we apply Eq, (36). If the conditions in Eqs. (44) and (46) are satisfied, the transition probability $p(t, \tilde{x}'|\tilde{x})$ assumes the form of [20]

$$p(t, \tilde{x}'|\tilde{x}) = \sqrt{\frac{p_{st}(t, \tilde{x}'|\tilde{x})}{p(0, \tilde{x}'|\tilde{x})}} \sum_{n=0}^{\infty} \psi_n(t, \tilde{x}')\psi_n(0, \tilde{x}) e^{-\lambda_n t}, \qquad \lambda_n \geq 0. \tag{47}$$

Because $p(\tilde{x}') \coloneqq p_{st}(t, \tilde{x}'|\tilde{x})$ is the integrand term of Eq. (43) for $\lambda_0 = 0$, for some constant $C$, we obtain

$$|p(t, \tilde{x}'|\tilde{x}) - p(\tilde{x}')| \leq \left|C \sum_{n=1}^{\infty} \psi_n(t, \tilde{x}')\psi_n(0, \tilde{x}) e^{-\lambda_n t}\right|$$

$$\leq C e^{-\lambda t}, \tag{48}$$

where $\lambda$ is a positive constant. This relation implies that the transition probability $p(t, \tilde{x}'|\tilde{x})$ convergence to an invariant probability measure $\rho_{x,y;\kappa,\mu}(\coloneqq p(\tilde{x}'))$ although the existence of such measure should be verified. The stationarity of tip of the curve is the necessary condition for ergodicity and the existence of $\rho_{x,y;\kappa,\mu}$ shows the mixing property of its dynamics. Considering the condition $\tilde{x} < 0$, the substitution $\tilde{x} \to -\tilde{x}$ in Eq. (46) yields the second condition in Eq. (40).

**Corollary 2.** *For $\kappa < 4$ and $t \to \infty$, the condition in Eq. (46) is transformed as*

$$-\frac{\sqrt{\kappa}}{2}\mu(\xi_t) \geq 0 \quad \text{for} \quad X_0^{\tilde{x}} = \tilde{x} > 0,$$

$$\frac{\sqrt{\kappa}}{2}\mu(\xi_t) \geq 0 \quad \text{for} \quad X_0^{\tilde{x}} = \tilde{x} < 0. \tag{49}$$

*Proof.* From Lemma 2, if $\delta > 2$, the Bessel flow $X_t^{\tilde{x}} \to \infty$ as $t \to \infty$. Regarding $X_t^{\tilde{x}}$ as the starting point of the BKE in Eq. (41), that is, $X_t^{\tilde{x}} \to \tilde{x}$, and considering the limit of $t \to \infty$, the relations in Eq. (49) are obtained.

**Remark 2.** *The tip of the curve generated by the standard SLE driven by the Wiener process has a stationary probability distribution for $\kappa < 4$ from the condition in Eq. (49) because $\mu(\xi_t) = 0$.* This result is consistent with that of Ref. 17.

The above-discussed ergodicity of the tip of the curve provides the equivalence between the ensemble average and the time average of $d(z \parallel \xi)$. This enables the expression of the free energy function using the complex Lyapunov exponent $\lambda^\xi$ defined in Eq. (23). The following theorem is the second main result of this study.

**Theorem 2.** *If the tip of the modified SLE curve is ergodic, the free energy $F(t)$ converges to a constant, i.e.,*

$$\lim_{t \to \infty} F(t) - F_{eq} = \frac{1}{2} k_B T \lambda^\xi, \tag{50}$$

*where, $\lambda^\xi$ is the complex Lyapunov exponent of the driving function.*

*Proof.* From the definition in Eq. (23), the complex Lyapunov exponent $\lambda^\xi$ measures how the initial infinitesimal separation $\delta z_0$ in the complex plane expands after time $t$. It denotes that

$$|\delta \xi_n| = |\delta z_0| e^{\lambda^\xi t}, \tag{51}$$

where $t = n\tau$, $\tau \in \mathbb{R}$, and we consider the limit of $\tau \to 0$. Based on the definition of $\lambda^\xi$, the following is obtained:

$$\lim_{\delta z \to 0} \frac{\delta \xi_n}{\delta z_0} = \prod_{i=0}^{n-1} \frac{d\omega_i(\xi_i)}{dz}. \tag{52}$$

Using Eqs. (6) and (52), $\lambda^\xi$ is expressed as

$$\lambda^\xi = \lim_{n\to\infty} \frac{1}{n} \ln \left| \prod_{i=0}^{n-1} \frac{d\omega_i(\xi_i)}{dz} \right| = \lim_{n\to\infty} \frac{1}{n} \sum_{i=0}^{n-1} \ln \left| \frac{d\xi_t}{dz} \right| = \lim_{n\to\infty} \frac{1}{n} \sum_{i=0}^{n-1} \ln \left| \frac{dg_t(z_t)}{dz} \right|. \quad (53)$$

Using Eq. (19), Eq. (53) leads to

$$\lambda^\xi = \lim_{n\to\infty} \frac{1}{n} \sum_{i=0}^{n-1} \frac{1}{2} d(z \parallel \xi). \quad (54)$$

If the ergodicity implied by Eq. (39) holds true, the right-hand side of Eq. (54) is replaced with the ensemble average of $\frac{1}{2} d(z \parallel \xi)$, that is,

$$\lambda^\xi = \frac{1}{2} D(z \parallel \xi), \quad \text{for} \quad t \to \infty. \quad (55)$$

Substituting Eq. (55) into Eq. (21), we obtain Eq. (50). If Eq. (24) holds true, this also indicates the convergence of the non-equilibrium free energy function $F(t)$ and the Gibbs entropy $S(t)$.

## V. CONCLUSION

In this study, the fundamental properties of the free energy corresponding to the tip of a modified SLE driven by the Langevin equation, are investigated. We prove, with some restrictions on the drift term of the driving function, the positivity of the time derivative of the free energy, which indicates the negative entropy production. Assuming the ergodicity implied by the stationarity of the dynamics of the tip of the modified SLE curve, we further demonstrated that the free energy convergences to a constant involving the complex Lyapunov exponent. These results are different from those of the previously reported *H*-theorem for the Fokker-Plank equation, while supporting the evidence that the modified SLE should be regarded as an open system. The present formula and results indicate that the increase of the free energy and negative entropy production are permitted in the theory of the Loewner evolution, providing a perspective to understanding the dissipative systems (e.g., morphogenetic systems or not conformally invariant turbulence). Thus, the dynamics generated by the modified SLE curve exhibit the *growth-induced stability*, in which the curve growth according to the time-dependent map consequently contributes the stability of the system itself like biological system.

**APPENDIX. CRITICAL DIMENSION OF BESSEL-TYPE PROCESS**

The Bessel-type process expressed by Eqs. (35) and (37) seems to have the critical dimensions, including the boundary that determines whether the downward flow of $X_t^{\tilde{x}}$ starting at $\tilde{x}$ returns to the origin or diverges to infinity in the real axis. For the usual SLE driven by the Wiener process, the associated Bessel process has a critical dimension at $\delta = \delta_c = 2$. We prove the existence of a critical dimension $\delta_c = 2$ for the modified SLE, using the conventional method for the SLE[18, 19] in the following.

Considering the Bessel flow expressed by (37) with the initial condition $X_0^{\tilde{x}} = \tilde{x}(>0)$, we define the first visiting time at $X_t^{\tilde{x}} = 0$ as [18,19]

$$T^{\tilde{x}} := \inf\{t > 0 : X_t^{\tilde{x}} = 0\}. \tag{A1}$$

Let us consider a continuous function

$$\phi(\tilde{x}) = \phi(\tilde{x}; \tilde{x}_1, \tilde{x}_2) = P[X_t^{\tilde{x}}(\sigma) = \tilde{x}_2], \quad 0 < \tilde{x}_1 < \tilde{x} < \tilde{x}_2 < \infty, \tag{A2}$$

where

$$\sigma := \inf\{t > 0 : X_t^{\tilde{x}} = \tilde{x}_1 \text{ or } X_t^{\tilde{x}} = \tilde{x}_2\}. \tag{A3}$$

Subsequently, we obtain the following relations.

$$\phi(\tilde{x}_1) = 0, \quad \phi(\tilde{x}_2) = 1. \tag{A4}$$

With the above definitions, let us consider the process $M_t := \mathrm{E}[\phi(X_t^{\tilde{x}}(\sigma))|\mathcal{F}_t] = \phi(X_{t\wedge\sigma}^{\tilde{x}})$, which is the conditional expectation of the function $\phi(X_t^{\tilde{x}}(\sigma))$ with respect to $\mathcal{F}_t$. From the definition of the filtration and the Markov property of the Ito diffusion, $M_t$ becomes a martingale.[18, 19] Applying the Ito formula to $M_t$ yields,

$$M_t = \phi(\tilde{x}) + \int_0^{t\wedge\sigma} \phi'(X_s^{\tilde{x}}) \left[dB_s + \frac{\sqrt{\delta-1}}{2}\mu(\xi_s) + \frac{\delta-1}{2}\frac{ds}{X_s^{\tilde{x}}}\right] + \int_0^{t\wedge\sigma} \frac{1}{2}\phi''(X_s^{\tilde{x}})\langle dB_s dB_s\rangle$$

$$= \phi(\tilde{x}) + \int_0^{t\wedge\sigma} \phi'(X_s^{\tilde{x}})dB_s + \int_0^{t\wedge\sigma} \frac{1}{2}\left\{\phi''(X_s^{\tilde{x}}) + \left[\sqrt{\delta-1}\mu(\xi_s) + \frac{\delta-1}{X_s^{\tilde{x}}}\right]\phi'(X_s^{\tilde{x}})\right\}ds. \quad (A5)$$

From the martingale property of $M_t$, the third term of the right-hand side of Eq. (A5) is zero. This results in the following differential equation:

$$\phi''(\tilde{x}) + \left[\sqrt{\delta-1}\mu(\xi_0) + \frac{\delta-1}{\tilde{x}}\right]\phi'(\tilde{x}) = 0, \qquad \tilde{x}_1 < \tilde{x} < \tilde{x}_2. \quad (A6)$$

For $\delta \neq 1$, one can easily observe that $\phi'(\tilde{x})$ satisfies

$$\phi'(\tilde{x}) = a\tilde{x}^{-(\delta-1)}\exp\left[-\sqrt{\delta-1}\int \mu(\xi_0)d\tilde{x}\right]. \quad (A7)$$

Here, $a$ is a constant. Defining

$$\Xi(\tilde{x}) := -\sqrt{\delta-1}\int \mu(\xi_0)d\tilde{x} = -\sqrt{\delta-1}\mu(\xi_0)\tilde{x} + C, \quad (A8)$$

where $C$ is a constant, $\phi'(\tilde{x})$ in Eq. (A7) becomes

$$\phi'(\tilde{x}) = a\tilde{x}^{-(\delta-1)}\exp\Xi(\tilde{x}). \quad (A9)$$

Integrating by parts and using $\phi(\tilde{x}_1) = 0$ in Eq. (A4), Eq. (A9) leads to

$$\phi(\tilde{x}) = a\left[\frac{\tilde{x}^{(2-\delta)}}{2-\delta}\exp\Xi(\tilde{x})\right]_{\tilde{x}_1}^{\tilde{x}} + a\sqrt{\delta-1}\int_{\tilde{x}_1}^{\tilde{x}} \frac{\mu(\xi_0)\tilde{x}^{(2-\delta)}}{2-\delta}\exp\Xi(\tilde{x})\,d\tilde{x}. \quad (A10)$$

Using $\phi(\tilde{x}_2) = 1$ in Eq. (A4), $1/a$ is expressed as

$$\frac{1}{a} = \frac{1}{2-\delta}\left[\tilde{x}^{(2-\delta)}\exp\Xi(\tilde{x})\right]_{\tilde{x}_1}^{\tilde{x}_2} + \frac{\sqrt{\delta-1}}{2-\delta}\int_{\tilde{x}_1}^{\tilde{x}_2}\mu(\xi_0)\tilde{x}^{(2-\delta)}\exp\Xi(\tilde{x})\,d\tilde{x}. \quad (A11)$$

Substituting Eq. (A11) into Eq. (A10), we obtain

$$\phi(\tilde{x}) = \frac{\left[\tilde{x}^{(2-\delta)}\exp\Xi(\tilde{x})\right]_{\tilde{x}_1}^{\tilde{x}} + \mu(\xi_0)\sqrt{\delta-1}\int_{\tilde{x}_1}^{\tilde{x}}\tilde{x}^{(2-\delta)}\exp\Xi(\tilde{x})\,d\tilde{x}}{\left[\tilde{x}^{(2-\delta)}\exp\Xi(\tilde{x})\right]_{\tilde{x}_1}^{\tilde{x}_2} + \mu(\xi_0)\sqrt{\delta-1}\int_{\tilde{x}_1}^{\tilde{x}_2}\tilde{x}^{(2-\delta)}\exp\Xi(\tilde{x})\,d\tilde{x}}. \quad (A12)$$

Repeating integrations by parts, Eq. (A12) is transformed into

$$\phi(\tilde{x}) = \frac{\left[\tilde{x}^{(2-\delta)}\exp\Xi(\tilde{x})\right]_{\tilde{x}_1}^{\tilde{x}} - \sum_{n=1}^{\infty}\left\{\left(-\mu(\xi_0)\sqrt{\delta-1}\right)^n\left[\tilde{x}^{(2-\delta+n)}\exp\Xi(\tilde{x})\right]_{\tilde{x}_1}^{\tilde{x}}\prod_{k=1}^{n}\frac{1}{2-\delta+k}\right\}}{\left[\tilde{x}^{(2-\delta)}\exp\Xi(\tilde{x})\right]_{\tilde{x}_1}^{\tilde{x}_2} - \sum_{n=1}^{\infty}\left\{\left(-\mu(\xi_0)\sqrt{\delta-1}\right)^n\left[\tilde{x}^{(2-\delta+n)}\exp\Xi(\tilde{x})\right]_{\tilde{x}_1}^{\tilde{x}_2}\prod_{k=1}^{n}\frac{1}{2-\delta+k}\right\}}. \quad (A13)$$

By calculating in a manner similar to the above for $\delta = 2$, $\phi(\tilde{x})$ is expressed as

$$\phi(\tilde{x}) = \frac{\int_{\tilde{x}_1}^{\tilde{x}} \frac{1}{\tilde{x}} \exp\Xi(\tilde{x})\, d\tilde{x}}{\int_{\tilde{x}_1}^{\tilde{x}_2} \frac{1}{\tilde{x}} \exp\Xi(\tilde{x})\, d\tilde{x}}. \tag{A14}$$

Repeating integrations by parts, Eq. (A14) is transformed into

$$\phi(\tilde{x}) = \frac{\left[(-\mu(\xi_0)\tilde{x})^{-1}\exp\Xi(\tilde{x})\right]_{\tilde{x}_1}^{\tilde{x}} + \sum_{n=1}^{\infty}\left\{\left[(\mu(\xi_0)\tilde{x})^{-(n+1)}\exp\Xi(\tilde{x})\right]_{\tilde{x}_1}^{\tilde{x}}(n+1)!\right\}}{\left[(-\mu(\xi_0)\tilde{x})^{-1}\exp\Xi(\tilde{x})\right]_{\tilde{x}_1}^{\tilde{x}_2} + \sum_{n=1}^{\infty}\left\{\left[(\mu(\xi_0)\tilde{x})^{-(n+1)}\exp\Xi(\tilde{x})\right]_{\tilde{x}_1}^{\tilde{x}_2}(n+1)!\right\}}. \tag{A15}$$

(i) For $\delta > 2$ and any fixed $\tilde{x}_2$, it can be observed that $\tilde{x}$ is bounded by $\tilde{x}_1 < \tilde{x} < \tilde{x}_2$. Noting $0 < \tilde{x} < \tilde{x}_2$ and using Eq. (A13) yields

$$\lim_{\tilde{x}_1 \to 0} \phi(\tilde{x}) = \lim_{\tilde{x}_1 \to 0} \frac{\left[\tilde{x}^{(2-\delta)}\exp\Xi(\tilde{x})\right]_{\tilde{x}_1}^{\tilde{x}} - \sum_{n=0}^{\lfloor \delta-2 \rfloor}\left\{b_n\left[-\tilde{x}_1^{(2-\delta+n)}\exp\Xi(\tilde{x}_1) + \tilde{x}^{(2-\delta+n)}\exp\Xi(\tilde{x})\right]\right\}}{\left[\tilde{x}^{(2-\delta)}\exp\Xi(\tilde{x})\right]_{\tilde{x}_1}^{\tilde{x}_2} - \sum_{n=0}^{\lfloor \delta-2 \rfloor}\left\{b_n\left[-\tilde{x}_1^{(2-\delta+n)}\exp\Xi(\tilde{x}_1) + \tilde{x}_2^{(2-\delta+n)}\exp\Xi(\tilde{x}_2)\right]\right\}}$$

$$= 1, \tag{A16}$$

where $b_n = \left(-\mu(\xi_0)\sqrt{\delta-1}\right)^n \prod_{k=1}^{n} \frac{1}{2-\delta+k}$ and $\lfloor \delta - 2 \rfloor$ denotes the integer part of $\delta - 2$.

(ii) For $\delta = 2$ and any fixed $\tilde{x}_2$, substituting $\tilde{x}_1 = 1/N$, Eq. (A15) yields

$$\phi(\tilde{x}, N) = \frac{\left[(-\mu(\xi_0)\tilde{x})^{-1}\exp\Xi(\tilde{x})\right]_{\tilde{x}_1}^{\tilde{x}} + \sum_{n=1}^{\infty}\left\{\left[(\mu(\xi_0)\tilde{x})^{-(n+1)}\exp\Xi(\tilde{x}) - \left(\frac{N}{\mu(\xi_0)}\right)^{(n+1)}\exp\Xi\left(\frac{1}{N}\right)\right](n+1)!\right\}}{\left[(-\mu(\xi_0)\tilde{x})^{-1}\exp\Xi(\tilde{x})\right]_{\tilde{x}_1}^{\tilde{x}_2} + \sum_{n=1}^{\infty}\left\{\left[(\mu(\xi_0)\tilde{x}_2)^{-(n+1)}\exp\Xi(\tilde{x}_2) - \left(\frac{N}{\mu(\xi_0)}\right)^{(n+1)}\exp\Xi\left(\frac{1}{N}\right)\right](n+1)!\right\}}. \tag{A17}$$

Subsequently, using Eq. (A17), we obtain

$$\lim_{\tilde{x}_1 \to 0} \phi(\tilde{x}) = \lim_{N \to \infty} \phi(\tilde{x}, N) = 1. \tag{A18}$$

(iii) For $1 \le \delta < 2$, using Eq. (A13), we obtain

$$\lim_{\tilde{x}_2 \to \infty} \lim_{\tilde{x}_1 \to 0} \phi(\tilde{x}) = 0. \tag{A19}$$

The Lemma 2 for the cases i) -iii) is proven by Eqs. (A16), (A18), and (A19).

Springer, 2016)

[20] H. Risken, *Fokker-Planck equation: methods of solution and applications. Springer series in synergetics*. (Springer, 1989)